\documentclass[aps, reprint, pre]{revtex4-1}
\usepackage{dcolumn}%
\usepackage{epsfig, graphics, amssymb, amsmath, color, bm}
\usepackage{epstopdf}

\usepackage{color}
\usepackage{amsmath}
\usepackage{color}
\usepackage{siunitx}
\usepackage{float}
\usepackage{soul}
\usepackage{pifont,wasysym}

\def\boldsymbol{\bm}

\usepackage[unicode=true,
  linktocpage,
  linkbordercolor={0.5 0.5 1},
  citebordercolor={0.5 1 0.5},
  linkcolor=blue, citecolor = blue, colorlinks=true]{hyperref}

\definecolor{dgreen}{rgb}{0.0, 0.4, 0.0}

\usepackage{lipsum} 

\begin{document}

\title{Swimming efficiency in a shear-thinning fluid}

\author{Herve Nganguia}
\thanks{These authors contributed equally.}
\affiliation{Department of Mechanical Engineering,
Santa Clara University, Santa Clara, California 95053, USA}
\author{Kyle Pietrzyk}
\thanks{These authors contributed equally.}
\affiliation{Department of Mechanical Engineering,
Santa Clara University, Santa Clara, California 95053, USA}
\author{On Shun Pak}
\email[Corresponding author: ]{opak@scu.edu}
\affiliation{Department of Mechanical Engineering,
Santa Clara University, Santa Clara, California 95053, USA}
\date{\today}

\begin{abstract}
Micro-organisms expend energy moving through complex media. While propulsion speed is an important property of locomotion, efficiency is another factor that may determine the swimming gait adopted by a micro-organism in order to locomote in an energetically favorable manner. The efficiency of swimming in a Newtonian fluid is well characterized for different biological and artificial swimmers. However, these swimmers often encounter biological fluids displaying shear-thinning viscosities. Little is known about how this nonlinear rheology influences the efficiency of locomotion. Does the shear-thinning rheology render swimming more efficient or less? How does the swimming efficiency depend on the propulsion mechanism of a swimmer and rheological properties of the surrounding shear-thinning fluid? In this work, we address these fundamental questions on the efficiency of locomotion in a shear-thinning fluid by considering the squirmer model as a general locomotion model to represent different types of swimmers. Our analysis reveals how the choice of surface velocity distribution on a squirmer may reduce or enhance the swimming efficiency. We determine optimal shear rates at which the swimming efficiency can be substantially enhanced compared with the Newtonian case. The non-trivial variations of swimming efficiency prompt questions on how micro-organisms may tune their swimming gaits to exploit the shear-thinning rheology. The findings also provide insights into how artificial swimmers should be designed to move through complex media efficiently.
\end{abstract}

\maketitle
\section{Introduction} \label{sec:Intro}

Motility of micro-organisms in fluids plays vital roles in diverse biological processes \cite{Fauci2006, Lauga2016}. The physics governing their locomotion at low Reynolds numbers in Newtonian fluids is relatively well understood \cite{lauga2009hydrodynamics}. However, micro-organisms often move through complex fluids displaying non-Newtonian fluid behaviors such as viscoelasticity and shear-thinning viscosity. The influences of these nonlinear rheological properties on biological locomotion at small scales and their implications on the design of artificial micro-swimmers are under active research. While extensive efforts have focused on the effects of viscoelasticity \cite{sznitman15, Gwynn_book}, much less is known about locomotion in shear-thinning fluids. Many biological fluids, including blood and respiratory and cervical mucus, display shear-thinning rheology, where the viscosity decreases nonlinearly with the shear rate \cite{bird1987dynamics}. Recent efforts have begun to seek answers to fundamental questions on locomotion in shear-thinning fluids  \cite{gooeyness}.

Shear-thinning rheology might be expected to enhance self-propulsion due to the reduction in viscous drag on the swimmer as the fluid becomes ``thinner" with actuations. The locomotion problem, however, embraces more complexity because the reduction in fluid viscosity could simultaneously reduce the propulsive thrust. \textcolor{black}{Asymptotic \cite{Rodrigo_lauga} and numerical \cite{montenegro2012modelling, Montenegro_shearthinning, Li2015_JFM} studies on undulatory swimmers as well as experiments on \textit{C.~elegans} \cite{Arratia_JFM, Park2016} found equal or greater swimming speeds in a shear-thinning fluid than in a Newtonian fluid. A recent experiment on helical propulsion \cite{Gomez2017} also observed enhanced speeds. On the other hand, the shear-thinning rheology was also shown to reduce the swimming speed in other scenarios \cite{montenegro2012modelling, Montenegro_shearthinning, Dasgupta2013, Datt2015_JFM, Riley2017}. These findings suggest that whether a swimmer displays a faster or slower swimming speed in a shear-thinning fluid largely depends on the class of swimmer \cite{montenegro2012modelling, Montenegro_shearthinning, qiu2014swimming} and details of its swimming gait \cite{Datt2015_JFM}.}

While propulsion speed is an important property of locomotion, a swimmer may also adjust its propulsion mechanism to reduce the energetic cost of moving through a medium at the expense of speed depending on the biological scenarios and environmental constraints. The concept of efficiency is often introduced in the analysis of locomotion.
 A swimming gait that maximizes the propulsion speed may not necessarily enhance the swimming efficiency.
\textcolor{black}{It is therefore biologically relevant to investigate how the properties of a swimmer and its surrounding medium influence the efficiency of locomotion. The classical definition of thermodynamic efficiency was proved difficult to apply in swimming at low Reynolds numbers \cite{childress2012}. Lighthill introduced the Froude efficiency, a concept coming from propeller theory, to characterize the efficiency of low-Reynolds-number swimmers \cite{lighthill1951, Lighthill1975}. The swimming efficiency is defined as $\eta = \mathcal{D} U / \mathcal{P}$, which compares the total power dissipation $\mathcal{P}$ in the fluid during swimming with a useful power output, defined as the power against the drag $\mathcal{D}$ in moving a rigid body of identical shape as the swimmer at the swimming speed $U$.
This standard definition has been widely adopted to characterize the efficiency of different low-Reynolds-number swimmers in Newtonian fluids \cite{Childress1981, stone_reciprocal, Chattopadhyay2006, Leshansky2007, Tam2007, Michelin2010, Ishimoto2014, Wiezel2016}. The efficiency of swimming in shear-thinning fluids, however, remains largely unexplored despite its biological relevance.}

Recently, some undulatory swimmers have been shown to dissipate less power ($\mathcal{P}$) during swimming in a shear-thinning fluid \cite{Rodrigo_lauga, Li2015_JFM, Gagnon2016_JFM}. Information on how shear-thinning rheology alters the useful power output ($\mathcal{D} U$) is still required to quantify their swimming efficiency. Even though these undulatory swimmers display equal or greater speeds ($U$), whether they generate more or less useful power output still depends on how drag ($\mathcal{D}$) is modified in the shear-thinning fluid. Since both the useful power output and the total power dissipation may be modified by shear-thinning rheology in non-trivial ways, it is difficult to predict the resulting effect on the swimming efficiency \textit{a priori}.
Much less is known about the power dissipation and efficiency of other types of swimmers in shear-thinning fluids. 
Here we ask the questions: Is swimming in a shear-thinning fluid more efficient or less than in a Newtonian fluid? How does the swimming efficiency depend on the propulsion mechanism of a swimmer (e.g.~pushers vs.~pullers) in a shear-thinning fluid? 
The answers to these fundamental questions provide insights into how micro-organisms and artificial micro-swimmers can move through biological media displaying shear-thinning rheology in energetically favorable ways.

In this work, we probe the answers to the above questions by considering the squirmer model \cite{lighthill1951, Blake1971} as a general locomotion model to represent different types of swimmers in a shear-thinning fluid. We explicitly calculate the swimming efficiency of a squirmer to reveal how shear-thinning rheology affects the efficiency of locomotion at low Reynolds numbers. 

The paper is organized as follows. In Sec.~\ref{sec:Formulation}, we formulate the problem by introducing the squirmer model (Sec.~\ref{sec:squirmer}) and governing equations for the shear-thinning fluid medium (Sec.~\ref{sec:ST}) before discussing the asymptotic limits considered in this work (Sec.~\ref{sec:Asymptotic}). In Sec.~\ref{sec:RT}, we employ a reciprocal theorem approach to calculate the power dissipation and the swimming efficiency of a squirmer in a shear-thinning fluid, bypassing detailed calculations of the non-Newtonian flow. The results are discussed in Sec.~\ref{sec:RD} before some concluding remarks in Sec.~\ref{sec:Conclusion}.

\section{Formulation} \label{sec:Formulation}
\subsection{The squirmer model} \label{sec:squirmer}
The squirmer model, first studied by Lighthill \cite{lighthill1951} and Blake \cite{Blake1971} to model the propulsion of ciliated protozoa, is arguably the simplest possible three-dimensional swimmer of finite size. The motion of beating cilia is represented as a distribution of velocities on the squirmer surface. For a steady spherical squirmer of radius $a$, the tangential, time-independent surface velocity distribution is decomposed into a series of the form  \cite{Pedley16}
\begin{align}
u_\theta (r=a, \theta) = \sum^\infty_{k=1}- \frac{2}{k(k+1)}B_k P^1_k(\cos\theta), \label{eqn:squirming}
\end{align}
where $P^1_k$ represents the associated Legendre function of the first kind, $\theta$ is the polar angle measured from the axis of symmetry, and the squirming modes $B_k$ can be related to Stokes flow singularity solutions. In a Newtonian fluid, only the $B_1$ mode (a source dipole) contributes to the propulsion speed $U_N = 2B_1/3$, and the $B_2$ mode (a force dipole) is the slowest decaying spatial mode that dominates the far-field velocity generated by a squirmer. Therefore, many studies considered model swimmers represented by only the first two modes of the expansion \cite{Pedley16}.

Although the squirmer model was developed originally for swimming ciliates (such as \textit{Volvox} \cite{Drescher2009_prl}), it has also gained popularity as a general locomotion model \cite{Magar2003, Ishikawa2008, wang2012, Michelin2013, Yazdi2015, Chisholm2016}. The parameters in the squirming modes can be adjusted to represent different types of swimmers, broadly categorized as pushers ($\alpha=B_2/B_1<0$), pullers ($\alpha>0$), and neutral squirmers ($\alpha=0$). A pusher, such as the bacterium \textit{Escherichia coli}, obtains its thrust from the rear part of the body. A puller, such as the  alga \textit{Chlamydomonas}, obtains its thrust from the front part. A neutral squirmer generates a surrounding flow corresponding to a source dipole.

 \subsection{Governing equations}\label{sec:ST}

The incompressible flow around a squirmer in a  shear-thinning fluid at low Reynolds number is governed by the continuity equation and Cauchy's equation of motion 
\begin{align}
\nabla\cdot \mathbf{u} &= 0, \\
\nabla\cdot\mathbf{T}  &= \mathbf{0},
\end{align}
where the stress tensor $\mathbf{T} = -p\mathbf{I} +\boldsymbol \tau$.
The constitutive equation for a shear-thinning fluid is given by the Carreau-Yasuda equation \citep{bird1987dynamics}: $\boldsymbol \tau = \left[ \eta_\infty + (\eta_0 - \eta_\infty)\left[ 1 + (\lambda_t|\dot{\boldsymbol \gamma}|)^2 \right]^{(n-1)/2} \right] \dot{\boldsymbol \gamma}$, where $\eta_0$ and $\eta_{\infty}$ represent the zero and infinite-shear rate viscosities respectively, and the strain rate tensor $\dot{\boldsymbol \gamma} = \nabla\mathbf{u} + (\nabla\mathbf{u})^T$ with its magnitude given by $|\dot{\boldsymbol \gamma}| = (\dot{\gamma}_{ij}\dot{\gamma}_{ij}/2)^{1/2}$. The power law index $n<1$ characterizes the degree of shear-thinning, and the relaxation time $\lambda_t$ sets the crossover strain rate at which the non-Newtonian behavior becomes significant. 
Rheological data of biological mucus can be well fitted by the Carreau-Yasuda model \citep{Rheology_mucus,Rodrigo_lauga}.

We non-dimensionalize lengths by the squirmer radius $a$, velocities by the first mode of actuation $B_1$, strain rates by $\omega = B_1/a$ and stresses by $\eta_0 \omega$. The dimensionless constitutive equation then takes the form
\begin{align}
\boldsymbol{\tau}^* = \left[ \beta + (1 - \beta)\left( 1 + Cu^2| \dot{\boldsymbol \gamma}^*|^2 \right)^{\frac{n-1}{2}} \right] \dot{\boldsymbol{\gamma}}^*, \label{eq:CarreauYasuda}
\end{align}
where the viscosity ratio $\beta = \eta_\infty/\eta_0 \in [0,1]$ and the Carreau number $Cu=\lambda_t\omega$, which compares the characteristic strain rate $\omega$ to the crossover strain rate $1/\lambda_t$ defined by the fluid relaxation time.

The deviatoric stress tensor $\boldsymbol \tau^*$ depends nonlinearly on the strain rate tensor $\dot{\boldsymbol \gamma}^*$. We first conduct asymptotic analyses to make analytical progress in the weakly nonlinear regime. Numerical simulations of the full problem later verify that the asymptotic results capture the essential behaviors of the system. The numerical simulations of the momentum equations at zero Reynolds number with the Carreau-Yasuda constitutive relation~Eq.~(\ref{eq:CarreauYasuda}) are implemented in the finite element method software COMSOL in a similar fashion reported in our previous work \citep{Datt2015_JFM}.

Hereafter, we drop the stars for simplicity and refer to only dimensionless variables unless otherwise stated.

\subsection{Asymptotic Analyses}\label{sec:Asymptotic}
The constitutive equation, Eq.~(\ref{eq:CarreauYasuda}), reduces to the Newtonian limit when $Cu = 0$ or $\beta = 1$. We investigate the weakly non-Newtonian behaviors by expanding Eq.~(\ref{eq:CarreauYasuda}) in the limits of small Carreau number ($\varepsilon = Cu^2 \ll 1$) or small deviation of the viscosity ratio from unity ($\varepsilon = 1-\beta \ll 1$) in regular perturbation series. In both cases, the constitutive equation with the leading-order non-Newtonian contribution takes the form
\begin{equation}\label{eqn:AsyExp} 
\boldsymbol \tau \sim \dot{\boldsymbol \gamma}_0 +\varepsilon \left(\dot{\boldsymbol\gamma}_1 + \mathbf{A} \right), 
\end{equation}
where for expansion in Carreau number ($\varepsilon = Cu^2 \ll 1$):
\begin{equation}\label{eqn:Amatrix_carreau}
\mathbf{A} = \frac{(1-\beta)(n-1)}{2} |\dot{\boldsymbol \gamma}_0|^2 \dot{\boldsymbol \gamma}_0,
\end{equation}
and for expansion in viscosity ratio ($\varepsilon = 1-\beta \ll 1$):
\begin{equation}\label{eqn:Amatrix_viscosity}
\mathbf{A} = \left[ -1 + (1 + Cu^2 |\dot{\boldsymbol \gamma}_0|^2)^{(n-1)/2} \right] \dot{\boldsymbol \gamma}_0.
\end{equation}

The non-Newtonian problem can be solved perturbatively to obtain the detailed flow surrounding the squirmer order by order. We instead bypass these detailed calculations via the reciprocal theorem \citep{happel2012low, stone_reciprocal,lauga2014locomotion,Corato2015_PRE} to calculate the power dissipation and the swimming efficiency of a squirmer in a shear-thinning fluid.

\section{A reciprocal theorem approach}\label{sec:RT}
To assess the efficiency of squirming in a shear-thinning fluid, it is necessary to consider the power $\mathcal{P}$  expended during the swimming process. Since the work done by the surface squirming motion is equal to the power dissipation in the fluid, we have
\begin{equation}\label{eqn:pwr_dissip1}
\mathcal{P} = - \int_S{\mathbf{T}\cdot\mathbf{n \cdot u}~{\rm d}S},
\end{equation}
where $S$ is the squirmer surface and $\mathbf{n}$ is the unit outward normal to $S$. In a Newtonian fluid, Stone and Samuel \cite{stone_reciprocal} applied the reciprocal theorem to obtain the swimming speed of a squirmer without knowledge of the surrounding flow. Although this approach cannot get as far regarding power dissipation, which involves gradients of the surrounding flow, bounds can still be set on the power dissipation and swimming efficiency with only knowledge of surface velocities \citep{stone_reciprocal}. 
The expression of power dissipation in a Newtonian fluid was obtained through detailed calculations by Lighthill \cite{lighthill1951} and Blake \cite{Blake1971} which, for a squirmer with only two modes, reads 
\begin{align}
\mathcal{P}_N = \frac{8 \pi (2+\alpha^2)}{3} \cdot
\end{align}

We calculate the leading-order correction to the power dissipation of a squirmer in a shear-thinning fluid in the asymptotic limits of $\varepsilon = Cu^2 \ll 1$ or $\varepsilon = 1-\beta \ll 1$ as 
\begin{equation}\label{eqn:pwr_dissip2}
\mathcal{P} \sim \mathcal{P}_N - \varepsilon \left( \int_S{\mathbf{T}_0\cdot\mathbf{n} \cdot \mathbf{u}_1~{\rm d}S} + \int_S{\mathbf{T}_1 \cdot \mathbf{n} \cdot \mathbf{u}_0~{\rm d}S} \right).
\end{equation}
The first integral in the bracket above
vanishes because $\mathbf{u}_1$ is a constant representing the first correction in propulsion speed on the squirmer surface and \textcolor{black}{hence $ \int_S{\mathbf{T}_0\cdot\mathbf{n} \cdot \mathbf{u}_1~{\rm d}S} = 0$} due to the force-free condition ($\int_S{\mathbf{T}_0\cdot\mathbf{n}~{\rm d}S}=\mathbf{0}$). The remaining integral
\begin{equation}\label{eqn:pwr_dissip2}
\mathcal{P} \sim \mathcal{P}_N - \varepsilon \int_S{\mathbf{T}_1 \cdot\mathbf{n} \cdot \mathbf{u}_0~{\rm d}S}
\end{equation}
involves the solution to the first-order non-Newtonian flow problem satisfying
\begin{align}
 \nabla\cdot \mathbf{u}_1 &= 0, \label{eqn:firstA}\\
\nabla\cdot\mathbf{T}_1  &= \mathbf{0}, \label{eqn:firstB}
\end{align}
where 
\begin{align}
\mathbf{T}_1 = -p_1 \mathbf{I} + \dot{\boldsymbol \gamma}_1 + \mathbf{A}. \label{eqn:T1}
\end{align}
Instead of solving Eqs.~(\ref{eqn:firstA}) and (\ref{eqn:firstB}), we obtain the correction to the Newtonian power dissipation via a reciprocal theorem approach proved effective in viscoelastic fluids \citep{Corato2015_PRE}. 

\begin{figure*}[!]
  \centerline{\includegraphics[width=0.85\textwidth]{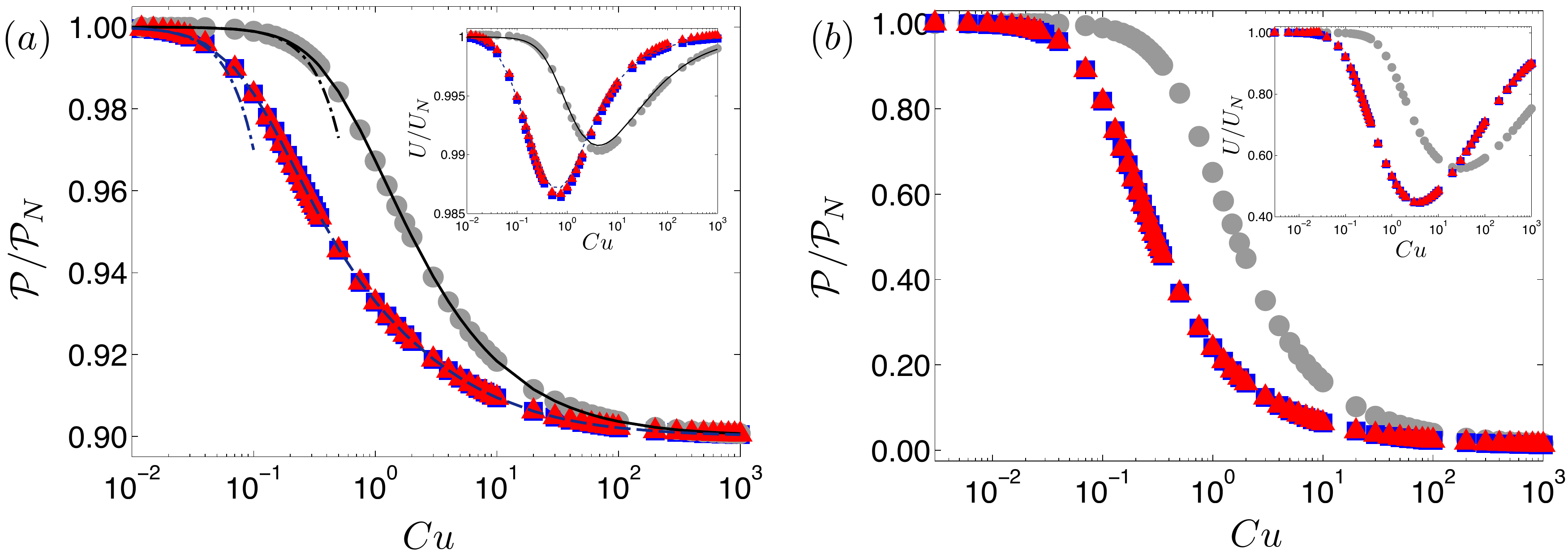}}
  \caption{(a) \textcolor{black}{Power dissipation in a shear-thinning fluid $\mathcal{P}$ (scaled by the corresponding Newtonian value $\mathcal{P}_N$)} for a neutral squirmer ($\alpha=0$, gray $\Circle$), puller ($\alpha=5$, blue $\square$), and pusher ($\alpha=-5$, red $\bigtriangleup$) by numerical simulations (symbols) agrees well with the asymptotic expansions in viscosity ratio for a neutral squirmer (solid line) and a pusher/puller (dashed line) at $\varepsilon = 1-\beta = 0.1$ and $n=0.25$. The variation in power dissipation at low shear rates is also well captured by the asymptotic expansion in Carreau number, $Cu$ \textcolor{black}{given by Eq.~\ref{eqn:Cu}} (dash-dotted lines). (b) Numerical computations of the power dissipation at $\varepsilon = 0.99$ and $n=0.25$. The insets show the corresponding swimming velocity $U$ compared with the Newtonian value $U_N$.
\label{fig1}}
\end{figure*}

To apply the reciprocal theorem, we consider the Newtonian squirming problem with known solutions \citep{lighthill1951, Blake1971} as the auxiliary problem, which satisfies
\begin{align}
\nabla\cdot \mathbf{u}_0 &= 0, \label{eqn:zeroA} \\
\nabla\cdot\mathbf{T}_0  &= \mathbf{0}, \label{eqn:zeroB}
\end{align}
where $\mathbf{T}_0= -p_0 \mathbf{I} + \dot{\boldsymbol \gamma}_0$. Taking the inner product of Eq.~(\ref{eqn:firstB}) with $\mathbf{u}_0$, minus the inner product of Eq.~(\ref{eqn:zeroB}) with $\mathbf{u}_1$, and integrating over the entire fluid volume $V$, we have $\int_V [\mathbf{u}_0 \cdot (\nabla \cdot \mathbf{T}_1) - \mathbf{u}_1 \cdot (\nabla \cdot \mathbf{T}_0)]~{\rm d}V = 0$. By vector calculus the integral can be rewritten as
\begin{align}
\int_V \nabla \cdot (\mathbf{u}_0 \cdot &\mathbf{T}_1 - \mathbf{u}_1 \cdot \mathbf{T}_0)~{\rm d}V  \nonumber \\
&= \int_V ( \nabla \mathbf{u}_0 : \mathbf{T}_1 - \nabla \mathbf{u}_1 : \mathbf{T}_0 )~{\rm d}V. \label{eqn:Re}
\end{align}
The left-hand side of Eq.~(\ref{eqn:Re}) can be converted into surface integrals by the divergence theorem and the right-hand side can be simplified by the constitutive equation Eq.~(\ref{eqn:T1}) as
\begin{align}
\int_S \mathbf{T}_1 \cdot \mathbf{n} \cdot \mathbf{u}_0~{\rm d}S -  \int_S \mathbf{T}_0 \cdot \mathbf{n} \cdot \mathbf{u}_1~{\rm d}S = - \int_V \mathbf{A}: \nabla \mathbf{u}_0~{\rm d}V.
\end{align}
Again, since $\mathbf{u}_1$ is a constant on the squirmer surface, the integral $\int_S \mathbf{T}_0 \cdot \mathbf{n} \cdot \mathbf{u}_1~{\rm d}S$ vanishes by the force-free condition and we obtain the result
\begin{align}
\int_S \mathbf{T}_1 \cdot \mathbf{n} \cdot \mathbf{u}_0~{\rm d}S  = - \int_V \mathbf{A}: \nabla \mathbf{u}_0~{\rm d}V. \label{eqn:T1A}
\end{align}
We substitute Eq.~(\ref{eqn:T1A}) into Eq.~(\ref{eqn:pwr_dissip2}) and obtain the first correction to the power dissipation due to the shear-thinning rheology for arbitrary distributions of surface velocity
\begin{align}
\mathcal{P} \sim \mathcal{P}_N + \varepsilon \int_V \mathbf{A}: \nabla \mathbf{u}_0~{\rm d}V, \label{eqn:ShearP}
\end{align}
which depends on only known solutions in Stokes flows \citep{lighthill1951,Blake1971}. Here $\mathbf{A}$ is given by Eq.~(\ref{eqn:Amatrix_carreau}) or Eq.~(\ref{eqn:Amatrix_viscosity}) depending on the asymptotic limit considered.

We use the standard definition of swimming efficiency introduced by Lighthill \cite{Lighthill1975} (see Introduction, Sec.~\ref{sec:Intro})
\begin{align}
\eta = \frac{\mathcal{D} U}{\mathcal{P}}, \label{eqn:Eff}
\end{align}
which compares the power dissipation $\mathcal{P}$ in the shear-thinning fluid due to the swimming motion [calculated by Eq.~(\ref{eqn:ShearP})] with the power required to drag a rigid sphere at the same swimming speed as the squirmer given by $\mathcal{D} U$. Here, $U$ denotes the swimming speed of the squirmer and $\mathcal{D}$ represents the force required to drag a rigid sphere at the swimming speed in the same fluid medium. Both quantities can be readily calculated in the two asymptotic limits considered in this work using the reciprocal theorem approach \citep{Datt2015_JFM}.

\begin{figure*}[!]
  \centerline{\includegraphics[width=0.85\textwidth]{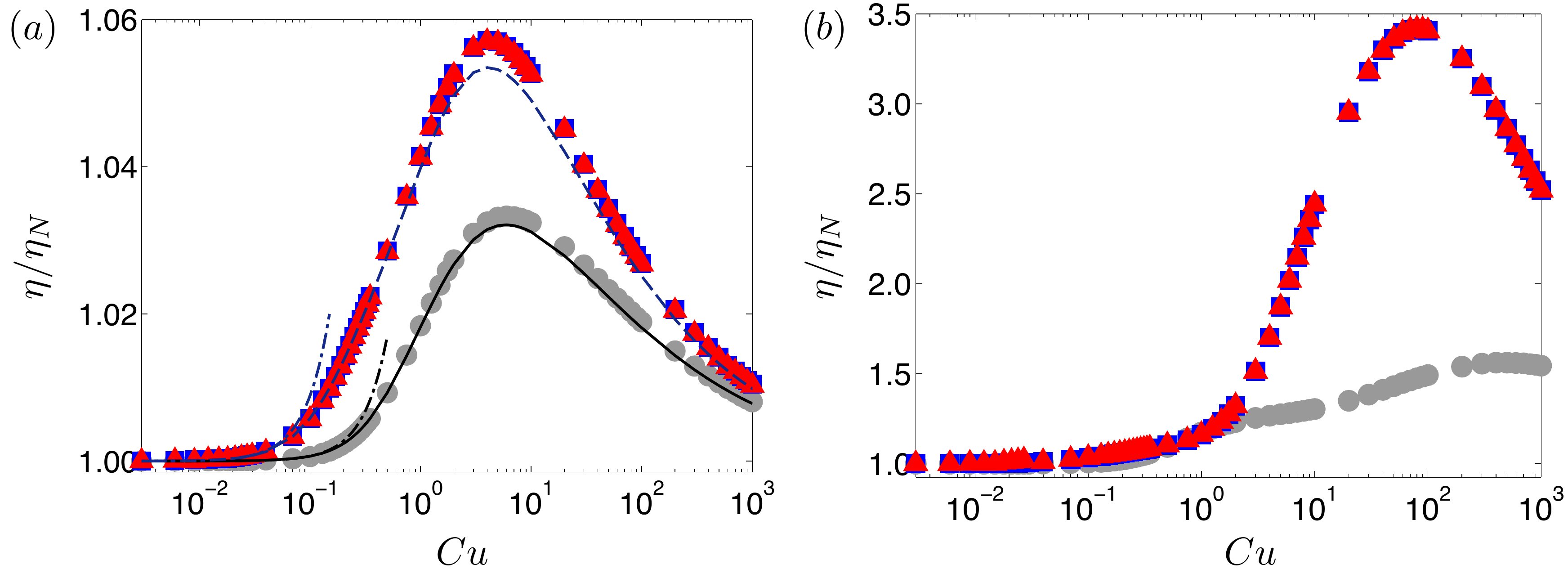}}
  \caption{(a) \textcolor{black}{Enhanced swimming efficiency in a shear-thinning fluid $\eta$ (compared with the Newtonian efficiency $\eta_N$)} for a neutral squirmer ($\alpha=0$, $\Circle$), puller ($\alpha=5$, $\square$), and pusher ($\alpha=-5$, $\bigtriangleup$) by numerical simulations (symbols) is well captured by the asymptotic expansions in viscosity ratio for a neutral squirmer (solid line) and a pusher/puller (dashed line) at $\varepsilon = 1-\beta = 0.1$ and $n=0.25$. The asymptotic expansion in Carreau number, $Cu$ \textcolor{black}{given by Eq.~\ref{eqn:STF_efficiency_2modes}} (dash-dotted lines) is effective in predicting the swimming efficiency at low shear rates. (b) Numerical computations of the power dissipation at $\varepsilon = 0.99$ and $n=0.25$. For a given viscosity ratio, an optimal $Cu$ maximizing the swimming efficiency of a swimmer exists. 
\label{fig2}}
\end{figure*}

\section{Results and Discussion} \label{sec:RD}
Here we employ the asymptotic results derived with the reciprocal theorem approach together with numerical simulations to explore how shear-thinning rheology affects the swimming efficiency of different types of swimmers. We first focus on the results for canonical two-mode squirmers in Sec.~\ref{sec:TwoMode} before investigating the effects of additional squirming modes in Sec.~\ref{sec:Others}.

\subsection{Two-mode squirmers} \label{sec:TwoMode}
In the squirmer model (see Sec.~\ref{sec:squirmer}), typically only the first two modes ($B_1$ and $B_2$) in Eq.~\ref{eqn:squirming} are retained to represent different types of swimmers \cite{Magar2003, Ishikawa2008, wang2012, Michelin2013, Yazdi2015, Chisholm2016}: pushers ($\alpha = B_2/B_1<0$), pullers ($\alpha >0$), and neutral squirmers ($\alpha = 0$). These two-mode squirmers were shown to always swim slower in a shear-thinning fluid than in a Newtonian fluid \cite{Datt2015_JFM} (see variations of propulsion speed $U$ reproduced as insets in Fig.~\ref{fig1}). Does the shear-thinning rheology render swimming of these two-mode squirmers more efficient or less? We first calculate the power dissipation during swimming in a shear-thinning fluid. 

\subsubsection{Power dissipation}

At low shear rates $\varepsilon=Cu^2\ll1$, \textcolor{black}{Eq.~(\ref{eqn:ShearP}) with Eq.~(\ref{eqn:Amatrix_carreau}) lead to an analytical expression for the power dissipation}
\begin{equation}
\frac{\mathcal{P}}{\mathcal{P}_N} \sim 1 + Cu^2(1-\beta)(n-1)\frac{C_1\left( C_2 + C_3\alpha^2 + \alpha^4 \right)}{2+\alpha^2}, \label{eqn:Cu}
\end{equation}
where $C_1=1.40$, $C_2=2.06$ and $C_3=5.75$ \textcolor{black}{are decimal numbers rounded up to represent lengthy fractions resulting from the analytical integration.}
Since the power law index $n < 1$ and viscosity ratio $\beta < 1$, Eq.~(\ref{eqn:Cu}) shows that shear-thinning rheology reduces the power dissipation of a squirmer [dash-dotted lines in Fig.~\ref{fig1}(a)]. At high strain rates as $Cu \rightarrow \infty$, from Eq.~(\ref{eqn:Amatrix_viscosity}) we have $\mathbf{A} \sim \dot{\boldsymbol \gamma}_0$ in which case the integral in Eq.~(\ref{eqn:ShearP}) reduces to 
$\int_V \mathbf{A}: \nabla \mathbf{u}_0~{\rm d}V \sim \int_V \dot{\boldsymbol \gamma}_0: \nabla \mathbf{u}_0~{\rm d}V = \mathcal{P}_N$. Therefore, the power dissipation asymptotes to $\mathcal{P} \sim (1-\varepsilon) \mathcal{P}_N = \beta \mathcal{P}_N$ as $Cu \rightarrow \infty$. That is, for a viscosity ratio $\beta=0.9$, the power $\mathcal{P} \sim 0.9 \mathcal{P}_N$ as shown in Fig.~\ref{fig1}(a). The total power dissipation decreases monotonically over the full range of $Cu$ for all two-mode squirmers  [black solid and blue dashed lines in Fig.~\ref{fig1}(a)] as revealed by results obtained with Eqs.~(\ref{eqn:ShearP}) and (\ref{eqn:Amatrix_viscosity}) in the asymptotic limit $\varepsilon = 1-\beta \ll1$. 
\textcolor{black}{The resulting integral is evaluated by quadrature because a closed form analytical expression is not available.}
These asymptotic results (represented by lines) in Fig.~\ref{fig1}(a) agree well with full numerical simulations (represented by symbols; see figure caption for details). In Fig.~\ref{fig1}(b), we run numerical experiments with $\varepsilon = 0.99$ and $n=0.25$ to emulate human cervical mucus \citep{Rheology_mucus, Rodrigo_lauga} and find similar trends. The monotonic decay in power dissipation for these squirmers is similar to the numerical and experimental results reported for undulatory swimmers \cite{Li2015_JFM, Gagnon2016_JFM}. 

We remark that  various types of non-Newtonian rheology (viscoelasticity and shear-thinning rheology) can affect the power dissipation of pushers vs.~pullers in different manners. The viscoelastic stress was shown to increase (decrease) the power dissipation for a pusher (puller) \citep{Corato2015_PRE}, however the shear-thinning rheology studied here reduces the power dissipation of a pusher and a puller indifferently as shown in Fig.~(\ref{fig1}) and by Eq.~(\ref{eqn:Cu}) for even powers of $\alpha$.

\subsubsection{Swimming efficiency}

Next, we use the power dissipation to calculate asymptotically the non-Newtonian correction to swimming efficiency defined in Eq.~(\ref{eqn:Eff}). At low shear rates $\varepsilon = Cu^2 \ll 1$, the swimming efficiency is given by
\begin{equation}\label{eqn:STF_efficiency_2modes}
\frac{\eta}{\eta_N} \sim 1 + Cu^2(1-\beta)(1-n)\frac{C_4 + C_5\alpha^2 + C_6\alpha^4}{2+\alpha^2},
\end{equation}
where $\eta_N = 1/(2+\alpha^2)$ represents the Newtonian swimming efficiency of a two-mode squirmer, $C_4=1.82$, $C_5=5.32$, and $C_6=0.30$.
The non-Newtonian correction due to shear-thinning rheology is strictly positive for arbitrary $\alpha$, meaning that all two-mode squirmers (neutral squirmers, pushers and pullers) display enhanced swimming efficiency at low shear rates.

At higher shear rates, the swimming efficiency varies non-monotonically as a function of $Cu$ in both the asymptotic [Fig.~\ref{fig2}(a)] and biologically relevant [Fig.~\ref{fig2}(b)] regimes. The swimming efficiency in a shear-thinning fluid is systemically higher than that in a Newtonian fluid ($\eta/\eta_N \ge 1$) over the full range of $Cu$ for all two-mode squirmers. 
There exist optimal shear rates (or optimal $Cu$) at which the swimming efficiency is maximized.
The numerical experiments with parameters emulating human cervical mucus in Fig.~\ref{fig2}(b) reveal substantial (more than threefold) enhancement in locomotion efficiency. The existence of an optimal $Cu$ may influence a swimmer to select a specific actuation rate for its swimming gait. When operating at the optimal $Cu$, a swimmer effectively exploits the shear-thinning rheology to gain the most energetically efficient propulsion.

\begin{figure}[t]
  \centerline{\includegraphics[width=0.4\textwidth]{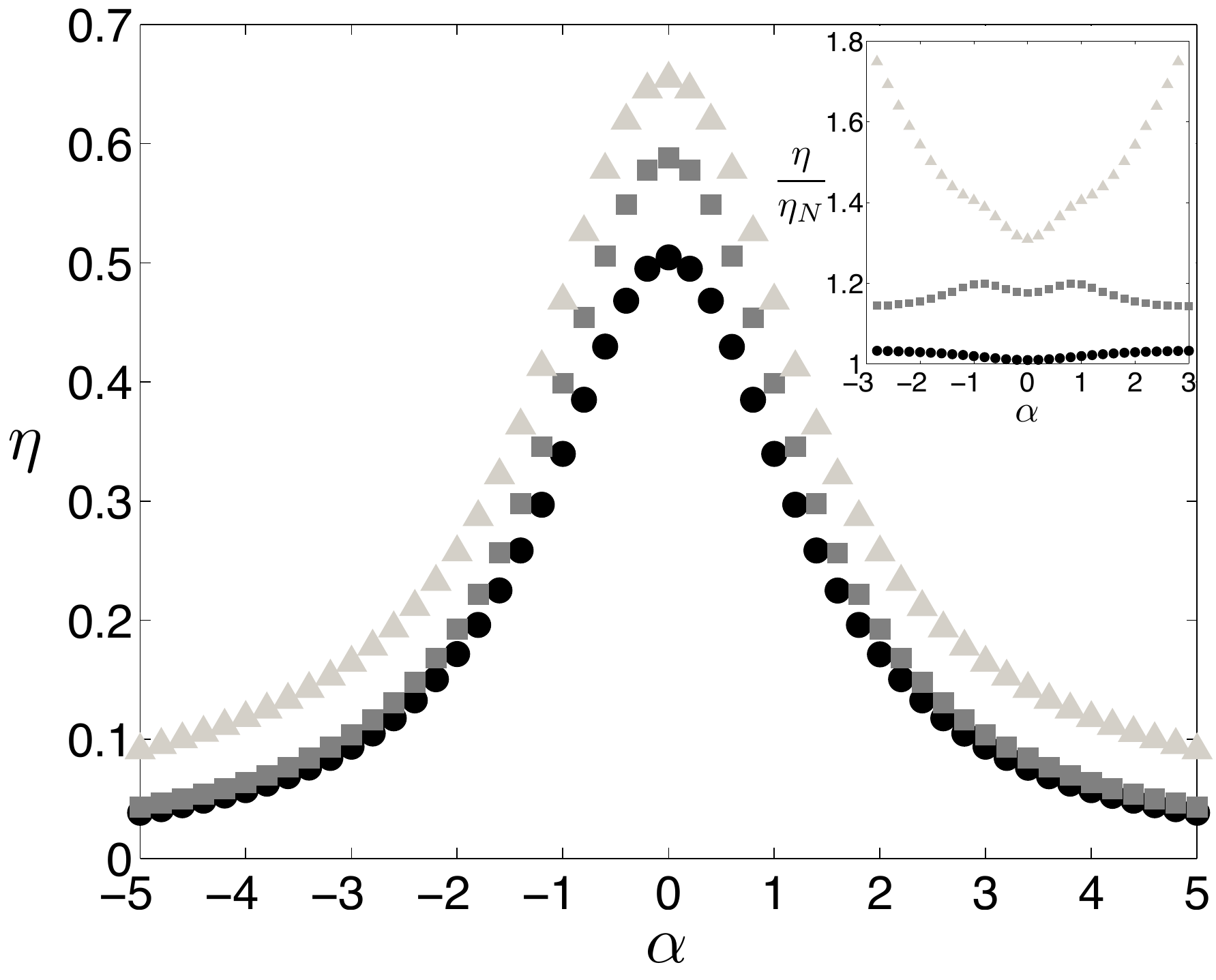}}
  \caption{\textcolor{black}{Swimming efficiency $\eta$ as a function of $\alpha=B_2/B_1$ at different values of $Cu$ with $\varepsilon = 1-\beta = 0.99$ and $n=0.25$ in a shear-thinning fluid: $Cu=0.1$ (black $\Circle$), $Cu=1$ (gray $\square$), and $Cu=10$ (light gray $\bigtriangleup$). The inset shows the enhancement in efficiency relative to the Newtonian value $\eta/\eta_N$ as a function of $\alpha$, where $\eta_N = 1/(2+\alpha^2)$}.
\label{EffAlpha}}
\end{figure}

\textcolor{black}{Next we investigate the variation of swimming efficiency as a function of the two squirming modes $\alpha = B_2/B_1$ at different values of $Cu$. We note that swimming efficiency in the Newtonian case $\eta_N = 1/(2+\alpha^2)$ is maximized with $\alpha=0$ (a neutral squirmer). In a shear-thinning fluid, larger percentages of efficiency enhancement relative to the Newtonian case ($\eta/\eta_N$) are observed for pushers and pullers ($\alpha \neq 0$, inset of Fig.~\ref{EffAlpha}). However, 
a neutral squirmer ($\alpha =0$) still maximizes the swimming efficiency in a shear-thinning fluid at different values of $Cu$ as shown in Fig.~\ref{EffAlpha}. The corresponding propulsion speeds at these maximum efficiencies are $U/U_N \approx 0.998$ (for $Cu=0.1$), $U/U_N \approx 0.886$ (for $Cu=1$), and $U/U_N \approx 0.588$ (for $Cu=10$). These results suggest that swimmers maximizing efficiency can still maintain considerable swimming speeds in a shear-thinning fluid.}

Overall, the above results demonstrate that although two-mode squirmers swim slower in a shear-thinning fluid than in a Newtonian fluid for all $Cu$ (Fig.~\ref{fig1} insets), they generally gain swimming efficiency in return. We also remark that the non-Newtonian correction to efficiency is even in $\alpha$ as shown in Eq.~(\ref{eqn:STF_efficiency_2modes}), which means that the shear-thinning rheology has exactly the same effect on the swimming efficiency of a pusher and a puller, again in contrast to the influence of viscoelasticity \citep{Corato2015_PRE}.

 \begin{figure*}[!]
   \centerline{\includegraphics[width=1.\textwidth]{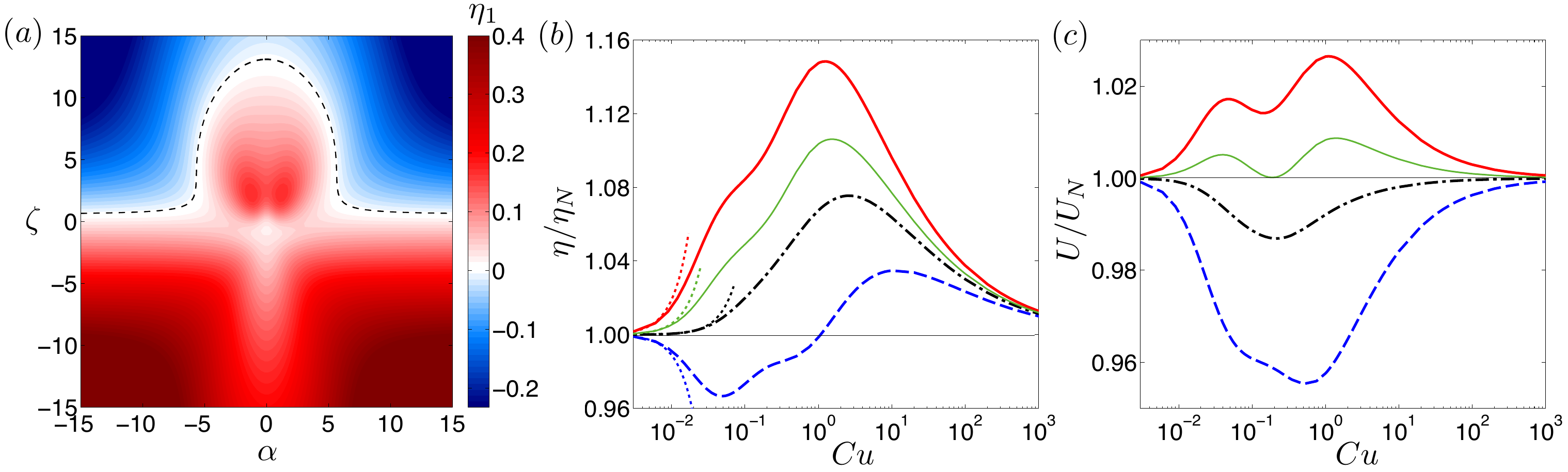}}
  \caption{ 
 \textcolor{black}{Swimming can be more efficient or less in a shear-thinning fluid, depending on the details of the swimming gait.} In (a), swimming efficiency at small Carreau number ($\eta \sim \eta_N + Cu^2 \eta_1$) given by Eq.~(\ref{eqn:STF_efficiency_3modes}) predicts instances of more efficient ($\eta_1>0$, the region colored in red) and less efficient ($\eta_1<0$, the region colored in blue) swimming depending on squirming modes represented by $\alpha = B_2/B_1$ and $\zeta = B_3/B_1$. The level set curve (dashed line) separates the two regions. For comparison, the efficiency and swimming speed of a two-mode squirmer ($\alpha =15$ and $\zeta=0$) are shown as black dash-dotted lines in (b) and (c) respectively. With only the first two modes, the swimmer gains efficiency but loses swimming speed. The addition of a third mode can either enhance or degrade the swimming performance depending on the choice of $\zeta$. For illustration, a swimmer chosen above the level set curve ($\alpha = 15$ and $\zeta=15$) displays in (b) less efficient swimming at small $Cu$ and in (c) slower swimming speed for all $Cu$, represented by the blue dashed lines; nevertheless, the swimmer can gain efficiency above the Newtonian limit at large $Cu$ in (b). \textcolor{black}{For the case $\alpha=15$, we determine a threshold value of $\zeta_c = -5.4$ below which both the efficiency and swimming speed of the squirmer are enhanced for all values of $Cu$, represented by thin green solid lines in (b) and (c).}  To further illustrate, a swimmer below the threshold value ($\alpha = 15$ and $\zeta=-15$) is chosen as another example to demonstrate that swimming in a shear-thinning fluid can be both (b) more efficient and (c) faster than the Newtonian case, represented by the red solid lines. The asymptotic expansion in $Cu$, Eq.~(\ref{eqn:STF_efficiency_3modes}), is effective in predicting the swimming efficiency at small $Cu$ (dotted lines) in (b).  
  \label{fig3}}
\end{figure*}
\subsection{Effects of other squirming modes} \label{sec:Others}
Squirming beyond the first two modes is not typically considered in a Newtonian analysis because other modes do not contribute to the swimming speed  \citep{lighthill1951, Blake1971}. \textcolor{black}{The presence of additional squirming modes hence simply reduces  the swimming efficiency in a Newtonian fluid \citep{stone_reciprocal}.} However, Datt \textit{et al.}~\cite{Datt2015_JFM} found that the shear-thinning rheology renders other squirming modes effective for propulsion. The addition of a $B_3$ mode alone can lead to non-trivial variations of swimming speed. Here we probe the effect of an additional squirming mode on the swimming efficiency in a shear-thinning fluid.
 
Using the same theoretical framework developed in previous sections, we calculate the swimming efficiency of a squirmer with a $B_3$ mode analytically and numerically. At small $Cu$, the swimming efficiency with the presence of a third squirming mode ($\zeta = B_3/B_1$) is given by
\begin{align}\label{eqn:STF_efficiency_3modes}
\frac{\eta}{\eta_N} \sim  1 &+ \frac{Cu^2(1-\beta)(1-n)}{4+2\alpha^2+ \zeta^2} \times  \bigg[ \textcolor{black}{2 C_4+2 C_5\alpha^2+2C_6\alpha^4}  \nonumber \\
&+\textcolor{black}{(C_{7}+C_{8}\alpha^2+C_{9}\alpha^4)\zeta  + (C_{10}+C_{11}\alpha^2)\zeta^2} \nonumber \\
&+\textcolor{black}{(C_{12}+C_{13}\alpha^2)\zeta^3+C_{14}\zeta^4+C_{15}\zeta^5 \bigg],}
\end{align}
where $\eta_N = 2/(4+ 2\alpha^2 + \zeta^2)$ represents the  Newtonian swimming efficiency of a three-mode squirmer, {\color{black} $C_{7}=4.48$, $C_{8}=12.84$, $C_{9}=-1.12$, $C_{10}=6.16$, $C_{11}=5.85$, $C_{12}=2.02$, $C_{13}=-0.68$, $C_{14}=0.60$, and $C_{15}=-0.06$}. The colormap in Fig.~\ref{fig3}(a) displays the non-Newtonian correction $\eta_1$ to swimming efficiency at low shear rates $\eta \sim \eta_N+Cu^2 \eta_1$ using the analytical expression, Eq.~(\ref{eqn:STF_efficiency_3modes}). 

In contrast to two-mode squirmers, which are systematically more efficient in a shear-thinning fluid than in a Newtonian fluid, the shear-thinning rheology can render swimming of a three-mode squirmer more efficient ($\eta_1>0$) or less ($\eta_1<0$) relative to the Newtonian case, depending on the choice of $\alpha$ and $\zeta$. The level set curves (dashed line) in the parameter space in Fig.~\ref{fig3}(a) separates the region of enhanced swimming efficiency (colored in red) from that of diminished swimming efficiency (colored in blue) at low shear rates. By adjusting the values of $\alpha$ and $\zeta$ in the swimming gait of a squirmer, we can construct swimmers that display non-trivial variations in both propulsion speed and swimming efficiency.

\textcolor{black}{For illustration, we first show in Fig.~\ref{fig3}(b) and (c) the efficiency and swimming speed of a two-mode squirmer $\alpha =15$ and $\zeta =0$ (black dash-dotted lines), where the two-mode squirmer gains efficiency at the expense of speed.} Based on Fig.~\ref{fig3}(a), we can construct a swimmer with $\alpha = 15$ and $\zeta=15$ in the blue region above the level set curve, which displays diminished swimming efficiency relative to the Newtonian efficiency at low $Cu$ (blue dash-dotted line) as shown in Fig.~\ref{fig3}(b). The variation of the swimming efficiency over the full range of $Cu$ however is non-monotonic (blue dashed line). Despite being less efficient than Newtonian swimming at low $Cu$, the same swimmer can gain efficiency above the Newtonian value at larger $Cu$ by increasing the actuation rate of the swimming gait. For this particular swimmer, its propulsion speed is systematically lower than the Newtonian speed for all $Cu$ [blue dashed line, Fig.~\ref{fig3}(c)]. Therefore, the addition of a third mode may harm both the propulsion speed and swimming efficiency compared with the Newtonian case for certain choices of parameter values.

\textcolor{black}{Nevertheless, the presence of a third mode also enables the design of a swimmer to propel both faster and more efficiently relative to the Newtonian case. For the case $\alpha=15$, by incrementally decreasing the value $\zeta$ below the level set curve in Fig.~\ref{fig3}(a), \textcolor{black}{we determine a threshold value of $\zeta_c = -5.4$ below which both the efficiency and swimming speed of the squirmer are enhanced for all values of $Cu$ [thin green solid lines in Figs.~\ref{fig3}(b) and (c)]}. To further illustrate, we construct a swimmer with $\alpha = 15$ and $\zeta=-15$, which displays both enhanced swimming efficiency [red solid line in Fig.~\ref{fig3}(b)] and propulsion speed [red solid line in Fig.~\ref{fig3}(c)] relative to the Newtonian case over the full range of $Cu$.}

These results illustrate the non-trivial variations in both propulsion speed and swimming efficiency due to the shear-thinning rheology. A swimmer may adjust the spatial (squirming modes) and temporal (actuation rate) details of its swimming gait for propulsion speed and/or swimming efficiency, depending on the biological scenarios and environmental constraints. The qualitative difference in the behaviors of two-mode and three-mode squirmers calls for caution in extending conclusions based on one type of swimmer to another.

\section{Conclusion} \label{sec:Conclusion}
Micro-organisms move through complex biological media that often display shear-thinning viscosity. This nonlinear rheology was shown to modify the propulsion speed of a swimmer in intriguing ways \cite{Rodrigo_lauga, montenegro2012modelling, Montenegro_shearthinning, Li2015_JFM, Arratia_JFM, Datt2015_JFM, Riley2017, qiu2014swimming, Gomez2017}. While the propulsion speed is an important property of locomotion, the efficiency is another factor that may determine the swimming gait adopted by a micro-organism in order to swim through complex media in energetically favorable ways. In this work, via the squirmer model, we reveal how swimming efficiency depends on the propulsion mechanism of a swimmer and the properties of its surrounding shear-thinning fluid.

Our analysis extends the classical results by Lighthill \cite{lighthill1951} and Blake \cite{Blake1971} in a Newtonian fluid to the case of a shear-thinning fluid. In the squirmer model, typically only the first two modes of surface velocity were considered in previous studies to represent neutral swimmers, pushers (e.g.~\textit{Escherichia coli}), and pullers (e.g.~\textit{Chlamydomonas}) \cite{Pedley16}. The shear-thinning rheology was shown to reduce the swimming speed of these two-mode squirmers \cite{Datt2015_JFM}. We show in this work that, although two-mode squirmers always swim slower in a shear-thinning fluid, they gain swimming efficiency in return. There exist optimal surface actuation rates (Carreau number) at which the swimming efficiency of these swimmers can be maximized. The optimal swimming efficiency can be substantially higher than the corresponding Newtonian efficiency. The enhancement in swimming efficiency provides a motivation for a swimmer to adjust its swimming gait and actuation rate to exploit the shear-thinning rheology for energetically efficient propulsion.

However, the above conclusion of enhanced swimming efficiency for two-mode squirmers should not be taken as \textit{a priori} for other types of swimmers. When an additional squirming mode is included on the squirmer surface, we show that swimming can become less efficient in a shear-thinning fluid than in a Newtonian fluid. Although the presence of the additional squirming mode may harm the swimming efficiency in some cases, we also demonstrate how the magnitude of this additional squirming mode can be adjusted to construct a swimmer that achieves the best of both worlds: swimming faster and more efficiently than in a Newtonian fluid.

The non-trivial variations of both the propulsion speed and swimming efficiency due to the nonlinear rheology suggest potentials in further optimization of locomotion performance otherwise impossible in a Newtonian fluid. \textcolor{black}{These findings also provide insights into how the actuation rate (modifying the Carreau number) and spatial distribution of actuation (modifying the relevant squirming modes) should be designed for artificial swimmers to move through shear-thinning media efficiently.}

\begin{acknowledgments}
O. S. Pak is grateful for the support from a Packard Junior Faculty Fellowship. The authors thank L. Zhu for his assistance in the numerical aspect of this work and C. Datt for helpful conversations.
\end{acknowledgments}

\bibliography{squirmer}

\end{document}